\documentclass[12pt,english,tikz]{article}

\usepackage[capposition=top]{float row}
\usepackage{latexsym}
\usepackage{amsfonts}
\usepackage{amsmath, amsthm, amssymb}
\usepackage{longtable}
\usepackage{caption}
\usepackage{subcaption}
\usepackage[round]{natbib}
\usepackage{multirow}

\usepackage{graphicx}
\usepackage{tikz}
\usetikzlibrary{trees}

\usepackage{dcolumn}

\usepackage{setspace}
\usepackage{color, colortbl}
\usepackage{graphicx}
\usepackage{multirow}
\usepackage{enumerate}
\usepackage{array}
\usepackage{catchfilebetweentags}
\usepackage{booktabs,calc}
\usepackage[]{amsmath}
\usepackage{float}
\usepackage{pdflscape}
\usepackage{rotating}

\newcommand{\specialLcell}[2][l]{%
  \begin{tabular}[#1]{@{}l@{}}#2\end{tabular}}  
\usepackage{tabularx}
\usepackage[colorlinks]{hyperref} 
\hypersetup{linkcolor=blue,urlcolor=blue, citecolor=black} 
\usepackage{verbatim}

\usepackage[all]{xy}
\usepackage{tikz}
\usetikzlibrary{arrows,shapes,backgrounds,calc}
\usepackage{onimage}
\definecolor{light-gray}{gray}{0.65}
\definecolor{dark-gray}{gray}{0.35}
\definecolor{LightCyan}{rgb}{0.88,1,1}
\definecolor{newblue}{rgb}{.031, .318, .612}
\usepackage[first=0,last=9]{lcg}
\usepackage{bbm}
\usepackage{titlesec}
\usepackage[skip=1pt]{caption}
\usepackage{subcaption}
\usepackage{amssymb}

\usepackage{soul}
\usepackage{setspace}
\usepackage{changepage}
\usepackage{chngcntr}
\usepackage{afterpage}
\usepackage{floatrow}

\usepackage{tocloft}
\usepackage{appendix}
\usepackage{etoolbox}

\AtBeginEnvironment{subappendices}{%
  \counterwithin{figure}{section}
  \counterwithin{table}{section}
}
\makeatletter
\patchcmd{\@chapter}{\protect\numberline{\thechapter}#1}
{\@chapapp~\thechapter: #1}{}{}
\makeatother

\newcolumntype{L}{>{\centering\arraybackslash}m{6cm}}
\newcolumntype{M}{>{\centering\arraybackslash}m{3cm}}
\newcolumntype{K}{>{\centering\arraybackslash}m{1.25cm}}
\newcolumntype{A}{>{\centering\arraybackslash}m{1.2cm}}

\newcolumntype{J}{>{\centering\arraybackslash}m{2.1cm}}
\newcolumntype{I}{>{\centering\arraybackslash}m{.745cm}}
\newcolumntype{B}{>{\centering\arraybackslash}m{.75cm}}

\newcolumntype{R}{>{\raggedright\arraybackslash}m{2.5cm}}
\newcolumntype{S}{>{\raggedright\arraybackslash}m{11cm}}
\newcolumntype{T}{>{\raggedright\arraybackslash}p{3.7cm}}
\newcolumntype{U}{>{\raggedright\arraybackslash}p{6cm}}
\newcolumntype{W}{>{\raggedright\arraybackslash}p{2.5cm}}
\newcolumntype{X}{>{\raggedright\arraybackslash}p{1.25cm}}
\newcolumntype{V}{>{\raggedright\arraybackslash}p{9cm}}
\newcolumntype{.}{D{.}{.}{-1}}
\newcolumntype{,}{D{,}{,}{-1}}
\newcolumntype{d}[1]{D{.}{.}{#1}}
\usepackage{setspace}
\theoremstyle{definition}

\usepackage{tabu}

\renewenvironment{abstract}
 {\small
  \begin{center}
  \bfseries \abstractname\vspace{-.5em}\vspace{0pt}
  \end{center}
  \list{}{%
    \setlength{\leftmargin}{4mm}%
    \setlength{\rightmargin}{\leftmargin}%
  }%
  \item\relax}
 {\endlist}

\usepackage{tikz}
\usetikzlibrary{calc,decorations.pathmorphing,patterns,shapes}

\pgfdeclaredecoration{penciline}{initial}{
    \state{initial}[width=+\pgfdecoratedinputsegmentremainingdistance,auto corner on length=1mm,]{
        \pgfpathcurveto%
        {
            \pgfqpoint{\pgfdecoratedinputsegmentremainingdistance}
                            {\pgfdecorationsegmentamplitude}
        }
        {
        \pgfmathrand
        \pgfpointadd{\pgfqpoint{\pgfdecoratedinputsegmentremainingdistance}{0pt}}
                        {\pgfqpoint{-\pgfdecorationsegmentaspect\pgfdecoratedinputsegmentremainingdistance}%
                                        {\pgfmathresult\pgfdecorationsegmentamplitude}
                        }
        }
        {
        \pgfpointadd{\pgfpointdecoratedinputsegmentlast}{\pgfpoint{0.5pt}{1.5pt}}
        }
    }
    \state{final}{}
}

\pgfdeclaredecoration{free hand}{start}
{
  \state{start}[width = +0pt,
                next state=step,
                persistent precomputation = \pgfdecoratepathhascornerstrue]{}
  \state{step}[auto end on length    = 3pt,
               auto corner on length = 3pt,               
               width=+2pt]
  {
    \pgfpathlineto{
      \pgfpointadd
      {\pgfpoint{2pt}{0pt}}
      {\pgfpoint{rand*0.15pt}{rand*0.15pt}}
    }
  }
  \state{final}
  {}
}
 \tikzset{free hand/.style={
    decorate,
    decoration={free hand}
    }
 } 
\def\freedraw#1;{\draw[free hand] #1;}

\usepackage{minitoc}
\usepackage{endnotes}

\usepackage[bottom]{footmisc} 

\usepackage{setspace}
\usepackage{titlesec}
\titleformat{\section}{\normalfont\Large\bfseries}{\thesection}{1em}{\setstretch{0.1}}
\titleformat{\subsection}{\normalfont\large\bfseries}{\thesection}{1em}{\setstretch{0.1}}

\RequirePackage{etex} 
\newcommand{\ajps}{0} 
\newcommand{\psrm}{0} 

\if\ajps1
\usepackage[margin=1in]{geometry}
\fi

\if\ajps0
\usepackage[margin=1in]{geometry}
\fi

\if\psrm1
\usepackage[right=3.5cm, left=3.5cm, top=3.5cm, bottom=3.5cm]{geometry}
\fi

\title{Political Conflict and Economic Growth in Post-Independence Venezuela}

\if\ajps0
\author{Dorothy Kronick\thanks{Assistant Professor of Public Policy, University of California, Berkeley.} \and Francisco Rodríguez\thanks{Rice Family Professor of the Practice of International and Public Affairs at the University of Denver’s Josef Korbel School of International Studies.}}
\fi

\date{\today}

\doparttoc
\dopartlof
\dopartlot
\faketableofcontents
\fakelistoffigures
\fakelistoftables

\begin{document}



\maketitle
\vspace{.25cm}
\vspace{0cm}

\thispagestyle{empty}

\begin{abstract}
Venezuela has suffered three economic catastrophes since independence: one each in the nineteenth, twentieth, and twenty-first centuries. Prominent explanations for this trilogy point to the interaction of class conflict and resource dependence. We turn attention to intra-class conflict, arguing that the most destructive policy choices stemmed not from the rich defending themselves against the masses but rather from pitched battles among elites. Others posit that Venezuelan political institutions failed to sustain growth because they were insufficiently inclusive; we suggest in addition that they inadequately mediated intra-elite conflict.
\end{abstract}
\begin{center}
\end{center}




\setcounter{page}{1}

\linespread{1.2}\selectfont

\newpage

The Venezuelan economy has suffered three catastrophes since independence: a long stagnation in the mid-nineteenth century, a devastating recession in the late twentieth century, and, in the second decade of this century, a record-setting collapse. In each of these three periods, the Venezuelan economy performed far worse than other countries in Latin America and far worse than other countries with similar exports.

It is tempting to explain these episodes as the predictable outcome of natural resource dependence. As the prices of Venezuelan exports rose and fell, so too did Venezuela’s economy. And indeed, each of Venezuela's major growth collapses can be traced in part to negative terms-of-trade shocks. But Venezuela is more than a simple case of the commodity lottery \citep{bulmer2003economic}: its growth contractions are much deeper than those of many countries that faced similar external conditions. 

Many scholars have argued that negative shocks do not inevitably trigger growth collapses; rather, negative shocks trigger growth collapses when domestic political institutions fail to mediate disagreements over who should bear the costs \citep[e.g.][]{rodrik1999did}. In bringing this potent argument to the Venezuelan case, much of the literature has focused on class conflict. Venezuela, the story goes, never adopted inclusive political institutions, even during the country’s democratic period \citep[e.g.][]{cordova1963estructura}; as a result, Venezuela ``languishes under extractive [economic] institutions’’ \citep[][p.\ 460]{robinson2012nations} that enable the rich to profit at the expense of the poor. Deterioration in the price of oil---Venezuela's principal export---exacerbates tension between the haves and the have-nots \citep{dunning2008crude}. In this view, the pathologies of natural resource dependence (such as the failure to diversify \citep{hausmann2014did} or the temptation to overspend during boom years \citep{karl1997paradox,edwards1991macroeconomic} ultimately stem from institutional arrangements designed to protect rent-seeking elites against competition and redistribution.

We argue that this focus on class conflict obscures the critical role of intra-class conflict in driving Venezuela’s economic performance \citep{kronick2021backsliding,rodriguez2021a}.\footnote{Our discussion also relates to a much broader literature on the links between economic growth and conflict \citep[e.g.][]{alesina1994distributive,benhabib1996social,collier1999economic,collier2003breaking,miguel2004economic}.} Venezuela’s nineteenth-century stagnation stemmed not so much from a conflict between elites and the masses but from a war between landowners and financiers that impeded the establishment of effective property rights. Only the definitive sidelining of the landowners paved the way for the centralization of state finances in the 1870s and, thereby, a resumption of economic growth, as we argue in Section 1. Similarly, in the lost quarter century between 1977 and 1999, battles between new- and old-guard elites hamstrung the country’s ability to adjust to tumbling oil prices (Section 2). And to the extent that Venezuela’s twenty-first century collapse stems from political conflict, we argue, it is not that of a revolutionary vanguard defending the poor against the rich but rather one of two elite factions engaged in an economically costly struggle for power (Section 3).

Our objective is not to develop a fully specified theory of which political institutions successfully manage intra-elite conflict, or to outline the conditions under which those institutions arise. Nor do we mean to overstate the similarity of three clearly distinct episodes. Rather, we seek to provide an empirical discussion that highlights the role of intra-class conflict in Venezuela’s political and economic development, and to suggest that intra-class conflict merits additional attention in the study of other cases.


\section{The Lost 19th Century}
In the sixteenth and early seventeenth centuries Venezuela ranked among the poorest parts of the Americas, scraping by on cacao exports while other colonies grew rich on gold and silver mining. Even cacao wealth, such as it was, suffered a blow in the form of a devastating blight in the 1630s \citep[][p.\ 625]{ferry1981encomienda}. But Venezuela’s fortunes changed by the beginning of the nineteenth century. The most reliable estimates of GDP per capita in 1800 suggest that Venezuela was richer than Brazil, Colombia, Chile, and even Peru \citep{bolt2020maddison}. Anecdotal evidence supports this conjecture. Alexander Humboldt, for example, who visited Caracas in 1799, claimed that he saw there (and in Havana) societies with “the most European physiognomy” making him feel “closest to Cádiz and the United States than anywhere else in the New World” (\citealp[p.~84]{von1814voyage}).  He also observed that “consumption of beef is immense in this country,” estimating per capita consumption at more than seven times that of Paris.  Another chronicler of the eighteenth century claimed that “all persons without distinction of age nor sex consume beef at least three times a day” \citep[p.~76]{loveraHistoriaAlimentacionVenezuela1988}. 

This prosperity stemmed from growth in the agricultural sector. Taking advantage of trade with the growing British Caribbean colonies, Venezuela became a diversified agricultural export economy earning significant income from cacao, indigo, coffee, and cattle hides \citep{ferrigniVenezuelaSeriesHistoria1986,ferrigniCrisisRegimenEconomico1999}. Outside of the Andes, large landholdings were concentrated in the hands of a small number of families.  

Because Venezuela was not known to hold large mineable metal deposits or dense exploitable native populations, Spanish colonization efforts were concentrated elsewhere. Yet in contrast to other neglected colonies, Venezuela's location made it a prime site for contraband trade with non-Spanish Caribbean colonies, which surged towards the end of the eighteenth century. This trade fostered an assertive colonial elite who spearheaded the struggle for independence in northern South America, making the territory central to Spanish efforts to retake the colonies. 

The colonial institutions that guaranteed property rights for these landowners disintegrated during the prolonged war of independence. Between 1815 and 1819, the loyalist administration expropriated 312 haciendas (70\% of those registered in 1810).  As patriot forces recovered these lands, they were used to compensate the victorious armies. While original landholders recouped most of their holdings at the end of the 13-year war, an important share ended up in the hands of military leaders. José Antonio Páez, for example, the military general who would become Venezuela’s first president after independence from Gran Colombia (1830), was said to personally hold a monopoly over beef sales in Caracas \citep{castilloblomquist1987}. But instability of property rights did not end with independence. 

\paragraph{The Conservative–Liberal contest that defined nineteenth-century politics in most of Spanish America was especially vicious in Venezuela.} We argue that this fact is essential to understanding growth performance. 

At the heart of the conflict, in the Venezuelan case, were rules governing access to credit. The Páez government liberalized credit in the 1930s, lifting interest-rate caps and instituting protections for creditors, to the benefit of the trading-house owners who provided most of the credit in the economy. This group became known as the Conservatives (confusingly, because they favored more liberal economic policies; the name arose simply because they sought to conserve the Páez status quo). Landowners and other debtors—collectively known as Liberals—soon grew to oppose these policies, pressing for debt relief and legislation protecting them against asset seizures. The Liberals also employed redistributive and anti-Caracas-elite rhetoric, such that their coalition eventually encompassed an incipient artisan class as well as some peasants.

\begin{figure}[t!] 
\RawFloats
       \centering
       \captionsetup{width=1\textwidth}
       \caption{Four Great Recessions}\label{fig:GDP}
      \caption*{\footnotesize{The black line plots an estimate of GDP per capita in Venezuela from \citet{maddison1995monitoring}; scaled so that the 2011 figure matches nominal 2011 GDP, according to the Venezuelan central bank (at the market exchange rate). The gray line plots an estimate of GDP from \citet{baptista1997bases}.  The gray shaded regions mark the long stagnation of 1837--1870, the recessions of 1890--1911 and 1978--1990, and the depression of 2012--2020.}} \vspace{-.01cm}
 \label{fig:mainresults}
  \begin{tikzonimage}[width=1\textwidth, trim= 0cm 0cm 0cm .5cm, clip=true]{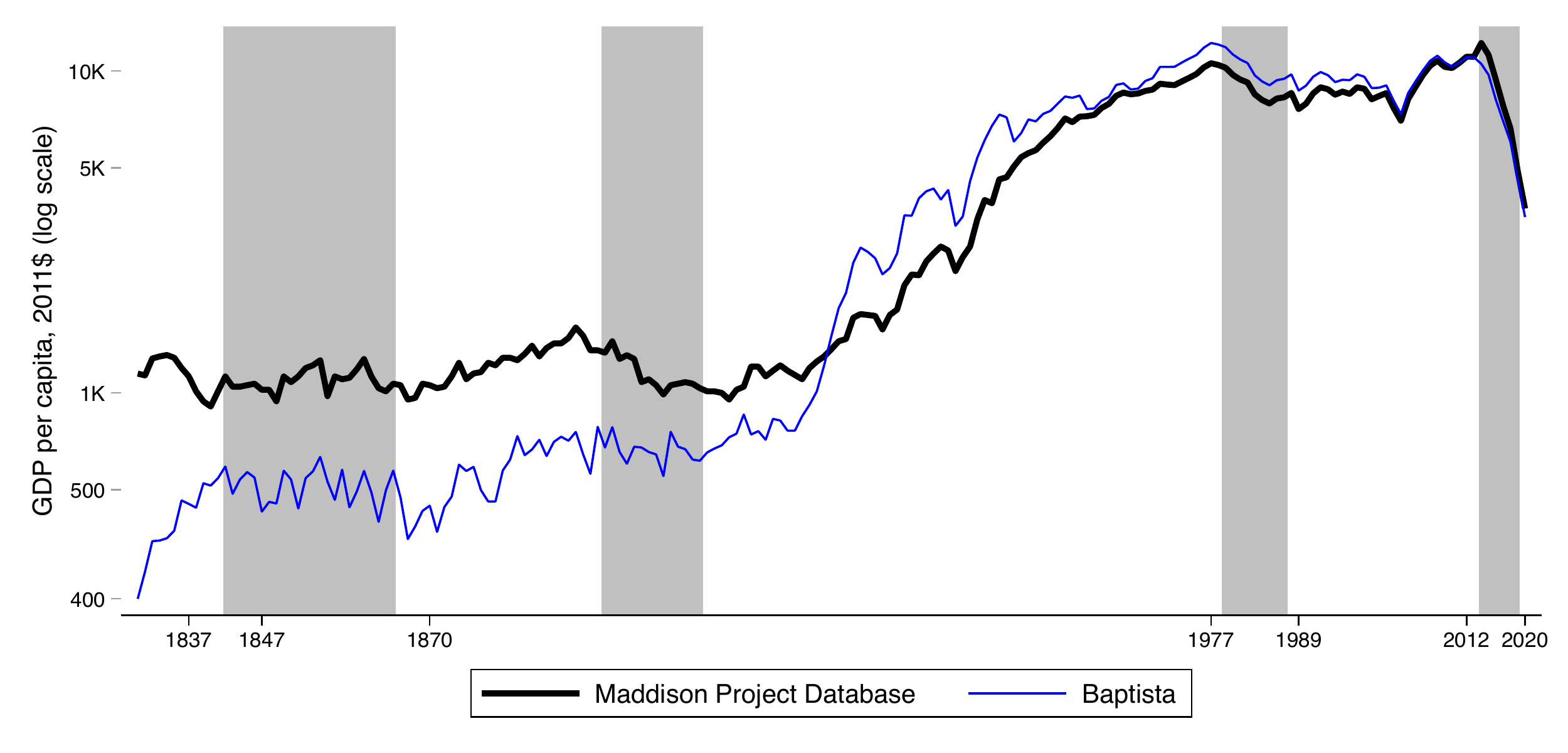}
 \end{tikzonimage}  \\[-1.25em]
\caption*{\scriptsize Sources: \citet{maddison1995monitoring}, \citet{baptista1997bases}, BCV}
 \end{figure}

Even before the destructive civil war of 1859–1863, which we address below, the Liberal–Conservative conflict hurt the agricultural sector. For one thing, instability in political leadership discouraged investment: every change in government implied risks for property owners who had sided with the prior dominant force. Such changes were frequent. Páez, for example, went from being head of state to leader of the main opposition faction at least three times in half a century. In this environment, it was difficult for hacendados to know whose good side they needed to be on \citep{britofigueroaProblemaTierraEsclavos1996}.

For another thing, Liberal governments faced a quandary typical of the ``game of bank bargains'' \citep{calomiris2014fragile}: their main constituents—indebted landowners—de\-manded debt relief whenever prices fell, but Liberal administrations could not afford to alienate either powerful international creditors or the local trading houses that underwrote public spending. Consider, for example, that a proposal to create an Institute of Territorial Credit that would financially assist distressed landholders was vetoed not only by Conservative governments but by a Liberal government as well. As a result, Liberal governments sometimes absorbed defaulted land debts, at significant cost to other public spending: in 1850, for example, Liberal President José Tadeo Monagas approved a debtor bailout that cost the equivalent of one year’s federal budget. In part as a result, the Monagas government could not control gangs of thieves who ran a contraband leather trade so active that it all but destroyed the cattle sector. 

Conflict between the Conservatives and the Liberals grew more heated throughout the 1840s and 1850s. Historian Reuban Zahlar has argued that there was no such thing as loyal opposition during this period; rather, both sides viewed the contest in Manichean terms \citep{zahler2013ambitious}. Liberal intellectual Tomás Lander wrote of ``two parties, Oligarchy [Conservatives] and Opposition [Liberals] $\ldots$ on one side wickedness and money, on the other side the boldness of [the Liberal newspaper], the justice that attends us and the innocence of the agriculturalists” (\citealp[p.~196]{zahler2013ambitious}). One of the founders of the Liberal party admitted frankly that many policies were defined simply in opposition to the Conservative position: ``Given that every revolt needs a rallying cry, and that the [Conservative] Constitutional Convention in Valencia rejected the label `federal’ for our constitution, let us take up that rallying cry ourselves—knowing, gentlemen, that had [the Conservatives] said `federal’ we would have rallied around centralism” \citep{gabaldon1988convencion}.

\begin{figure}[t!] 
\RawFloats
       \centering
       \captionsetup{width=1\textwidth}
       \caption{Terms of Trade, 1830--1920}\label{fig:TOT}
      \caption*{\footnotesize{This figure plots the ratio of an index of Venezuela's export prices to an index of its import prices, which began to improve in the late 1850s. The shaded regions correspond to the recessions marked in Figure \ref{fig:GDP}.}} \vspace{-.01cm}
 \label{fig:mainresults}
  \begin{tikzonimage}[width=1\textwidth, trim= 0cm 0cm 0cm .5cm, clip=true]{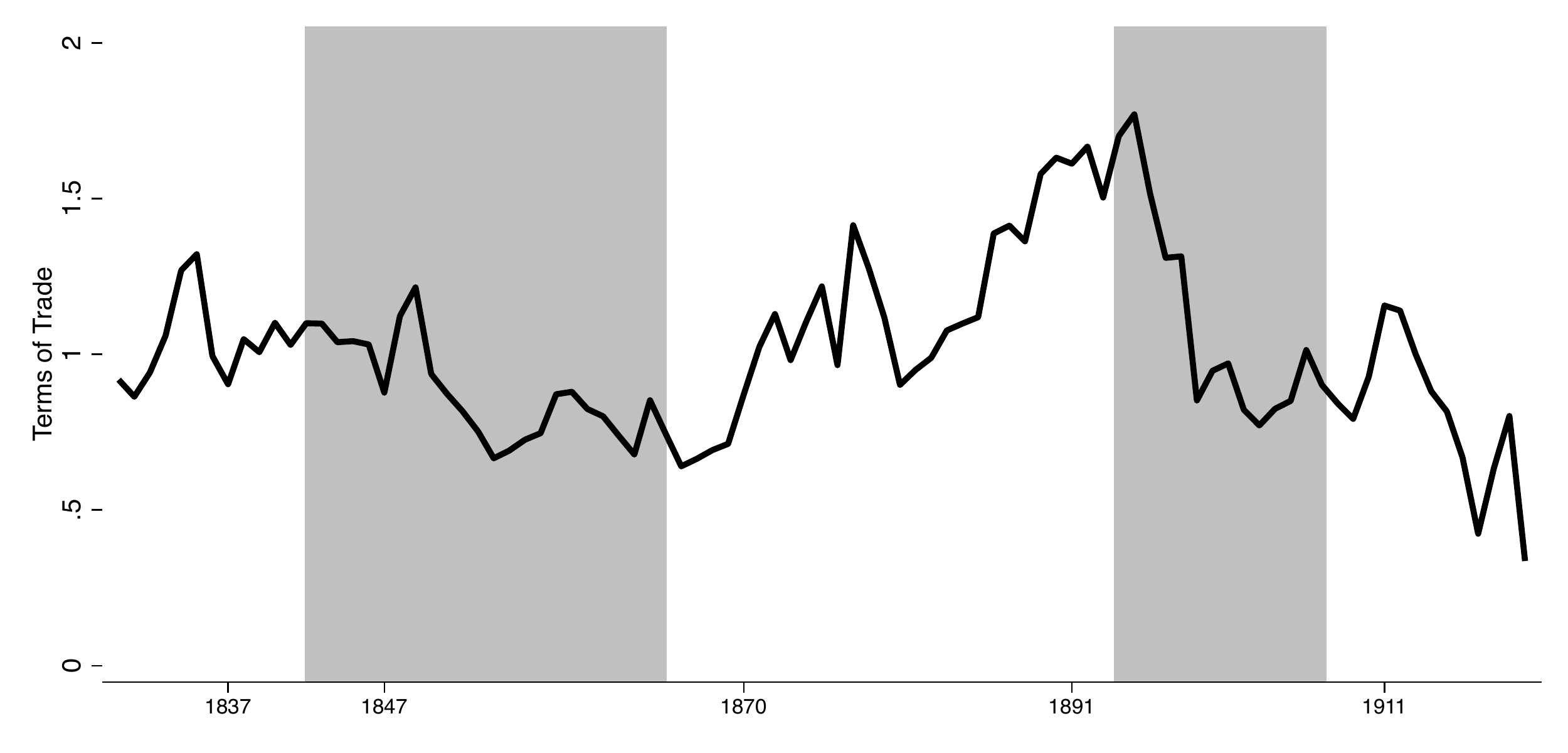}
 \end{tikzonimage}  \\[-1.25em]
\caption*{\scriptsize Source: \citet{baptista1997bases}}
 \end{figure}

This conflict soon devolved into war.  Discontent with the ``Liberal autocracy’’ (1847–1858) of José Tadeo Monagas and his brother José Gregorio drove a Liberal splinter faction to ally with Conservatives in overthrowing the government. The rivals fought intensely for five years (1858–1863). Certainly, the war derived in part from class conflict: the Liberal–Federal alliance was led by Ezequiel Zamora, who called for aggressive land reform and redistribution. “There will be neither rich nor poor, nor slaves nor owners, nor powerful nor discarded,” Zamora proclaimed, “but instead brothers who without lowering their brow will speak to each other as equals” \citep[p.~151]{frankel1976}. Ultimately, though, Zamora was assassinated and the conflict was settled between elite landholders and the financier class. 

The Federal War was extremely costly.  Estimates of deaths on the battlefield and from war-related illness range from 150 to 200 thousand people in a population of 1.8 million: that is, 8–11\% of the population \citep{vallenilla1997}. The economy contracted, though the magnitude of the wartime contraction varies across sources (Figure \ref{fig:GDP}): \citet{maddison1995monitoring} estimates put the 1858--70 contraction at a cumulative 4\% (following a period of fast growth), while \citet{baptista1997bases} estimates a much deeper 23\% contraction, about as much as the cumulative contraction of the U.S. economy during the Great Depression. Declining prices of Venezuelan exports (relative to imports) likely contributed to this recession (Figure \ref{fig:TOT}), but terms of trade actually increased 16.6\% between 1857 and 1870 \citep{baptista1997bases}, even as GDP continued to fall.  In short, the length of Venezuela’s nineteenth-century stagnation can only be explained by accounting for the destruction wrought by armed conflict.

%
%

\paragraph{Another way to understand the centrality of intra-elite conflict in driving growth} is to consider what happened when the conflict abated. Between 1870 and 1887, Antonio Guzmán Blanco largely (but temporarily) solved the mid-19th-century problems of endemic armed conflict and the consequent economic destruction, with the result that the economy grew quickly. In contrast, when Cipriano Castro (president from 1899--1908) antagonized financial elites toward the beginning of his presidency, another devastating civil war crippled the economy. Only the restoration of political stability positioned the economy to benefit from the beginning of oil production.  

Antonio Guzmán Blanco's pacifying innovation came after the Liberals---which is to say, the coalition of landowners that had traditionally opposed the Caracas-based financial class---won the Federal War. But when Liberal Guzmán Blanco took office, he did an about-face, ditching the support of (Liberal) landowners ``by surprise’’  \citep{stokesMandatesDemocracyNeoliberalism2001} to build an alliance with (Conservative) financiers. This move enabled a steady source of finance for the Venezuelan state.  Moreover, Guzmán designed a revenue-sharing agreement that temporarily brought regional caudillos under control of the central government. 

Customs collection was the key to centralization under Guzmán. Prior to his presidency, regional caudillos controlled their respective local trading houses and took a share of customs revenue. Guzmán instead centralized customs collection, offering the regional caudillos a fiscal transfer in return. This process began with partial privatization: in 1870, the Guzmán government created the Compañía de Crédito, a privately owned firm with minority government participation. The principal shareholders were the trading houses themselves, who were also large owners of government debt.  As the country’s importers, trading houses were well-placed to reduce customs corruption and contraband. In return for collecting customs, the Compañía could direct 85\% of revenue to paying debt obligations that the government owed to Compañía owners; the remaining 15\% went to the government. With the customs revenue as effective collateral, trading houses were then willing to issue new debt.

The regional caudillos accepted this arrangement because Guzmán offered them a constitutionally guaranteed fiscal transfer in return for ceding control of the customs houses. This transfer, which became known as the Situado Constitucional and survives to this day, provided the caudillos with a stable source of revenue.  Faced with a new conflict, Guzmán would offer the caudillos a choice: accept a share in higher levels of customs collection, or risk a confrontation with the central government, loyal caudillos, and the commercial–financial sector.

The Situado was not the only element of Guzmán’s coalition building.  Guzmán, the son of an influential Liberal politician, was not a member of any of Venezuela's regional groups and was therefore able to mediate conflict among the caudillos.  Rather than throwing the government’s weight behind those loyal to his party, Guzmán formally recognized dominant caudillos as presidents of their respective states. He also granted them control over electoral results by imposing a system of public, signed balloting; this system clearly favored those who held local military control. Moreover, Guzmán implemented a massive public works program directed through local promotion boards (Juntas de Fomento) in which regional caudillos and financiers were given seats \citep{quinteroSistemaPoliticoGuzmancista1994}.

Taken together, these measures allowed Guzmán to consolidate public finances and strengthen the central government, improving political and economic stability. But because the regional caudillos retained military power, the pax guzmánica was fragile: it rested in part on Guzmán’s web of personal alliances. Guzmán abandoned the presidency in 1887, apparently for health reasons, and moved to Paris, planning to run the presidency by proxy. But infighting erupted among rival Liberal factions and regional conflict returned. Guzmán’s abdication was followed by a twelve-year period of internecine wars during which Venezuela churned through seven presidents. The Situado survived, shoring up public finance, but in Guzmán’s absence it did not provide sufficient basis for political stability.

Guzmán presided over a period of political stability and economic growth.  Per capita incomes grew 43\% between his accession in 1870 and 1891, the last year before the Liberals were ousted by an armed insurrection (Figure \ref{fig:GDP}).  By contrast, the Venezuelan economy shrank by 18\% between 1835 and 1870 and by 37\% between 1891 and 1911, the two periods of political instability and armed conflict that preceded and followed the Long Guzmanato (the period between 1870 and the overthrow of the last of Guzmán's Liberal successors, Raimundo Andueza Palacios, in 1892).

While rising terms of trade (i.e., favorable prices for Venezuelan exports, relative to imports) certainly contributed to Guzmán’s capacity to forge a rent-sharing agreement with caudillos (Figure \ref{fig:TOT}), the economy began shrinking as the Long Guzmanato came to an end---even while terms of trade continued rising.  Remarkably, the descent into conflict after the collapse of the Long Guzmanato prevented Venezuela from participating in the region’s ``export age'' (as Bertola and Ocampo characterize it in their introductory chapter to this volume), only rejoining the period of outward-oriented growth with the emergence of the oil sector in the 1920s.  

These data support our thesis that Venezuela’s economy experienced growth when Guzmán quelled intra-elite conflict through the centralization of state finances.  When his architecture collapsed, intra-elite conflict and economic decline returned. The latter half of the nineteenth century might be viewed as a long civil war punctuated by two decades of stability under Guzmán's coalition among regional caudillos, the Liberal party, and the financial sector.

The presidency of Cipriano Castro (1899--1908) clearly illustrates the consequences of rekindling conflict with financial elites. When the powerful banker José Antonio Matos Tinoco (who, not incidentally, was Guzmán's brother-in-law) refused to lend more money to Castro's government, Castro responded not by negotiating but with a show of force: he publicly humiliated Matos and his associates, parading them through Caracas and throwing them in prison. Matos then granted the loan, but also marshaled troops for an attempt to overthrow Castro. The resulting civil war---the Revolución Libertadora, 1901--1903---took tens of thousands of lives and came at tremendous cost to the economy. Castro also refused to pay international debts, leading to the 1902 blockade of Venezuelan ports by Great Britain and Germany. 

Castro’s successor, Juan Vicente Gómez, rebuilt the alliance between the government and financiers. Gómez appointed Matos himself as foreign minister. And Gómez met weekly with Vicente Lecuna, president of the Caracas Chamber of Commerce and director of Banco de Venezuela; Lecuna went on to draft a monetary reform, at Gómez’s request. Financiers also supported the administrative reforms of Román Cárdenas, finance minister between 1913 and 1922. Cárdenas imposed unity of the treasury, which meant that all receipts and payments of the government would be centralized in the finance ministry. By 1930, Gómez proudly informed the country that Venezuela had paid all its external debt \citep{kornblithGestionFiscalCentralizacion1981}. 

Gómez also consolidated military power, finishing the process of centralization that Guzmán had started. Among the crucial military reforms was the creation of four traditional military divisions—infantry, cavalry, artillery and engineering—and the founding of a Military Academy for officer training. Gómez's predecessor, Cipriano Castro, had created the position of Inspector General of the Army, which integrated the political and administrative activities of the armed forces. And critically, the Castro administration made a massive purchase of repeat rifles, giving the army a definitive firepower advantage over regional caudillos \citep{ziemsEjercitoAlcanceNacional1993,rangelGomezAmoPoder1974}. The Andeans thereby created a modern national army with a centralized command system, packed with Andean loyalists.  Andean military commanders then ran Venezuela for all but three of the sixty years following Castro's ascent in 1899. 

By the second decade of the nineteenth century, then, Venezuela had a centralized state with a professionalized military and consolidated public finances.  Regional caudillos had been sidelined; Gómez triumphantly claimed the title “Pacifier of Venezuela.” This degree of state-building was by no means the norm in Latin America. Mexico’s Porfirio Díaz, for example, had scrapped a project to professionalize the Mexican military and was unable to concentrate fiscal collection in the hands of the central government \citep{hamnettConciseHistoryMexico2006,diaz-cayerosFederalismFiscalAuthority2006}—with the result that regional caudillos still held considerable power when Díaz was overthrown in 1911. 

This was the state that oversaw the advent and growth of the Venezuelan oil sector. Production began in 1914 and gained macroeconomic significance by the early 1920s. Growth was vertiginous. By 1929, Venezuela was the world’s largest oil exporter, and oil prices generally increased for the next half century. As a result, the Venezuelan economy grew quickly: GDP per capita increased by a factor of more than ten between 1920 and 1975 (Figure \ref{fig:GDP}), outperforming all other economies in Latin America. 

It is beyond the scope of this chapter to provide a comprehensive account of which policies fostered this growth and which policies hindered it. Instead, we focus on three decisions that are especially relevant for understanding the economic collapse of the 1980s.

First, Venezuelan governments did little to counter the effects of the exchange rate appreciation that comes with growing oil revenues. By and large, Venezuela embraced a more appreciated real exchange rate rather than trying to offset or slow appreciation, effectively discouraging growth in export sectors other than oil. This policy emerged in Venezuela’s response to the 1933 decision by U.S.\ President Franklin Delano Roosevelt to take the dollar off the gold standard.  While most Latin American nations reacted by devaluing their currencies—Argentina and Brazil had in fact done so before the United States—Venezuela did not, instead accepting a significant real exchange rate appreciation. This choice was not surprising given that the commercial–financier class had won the decades-long conflict with landowners: in other words, the sectors that would have advocated for a weaker exchange rate lacked political muscle.\footnote{Real appreciation is an increase in the real exchange rate, which is the price of domestic goods in terms of international goods, expressed in the same currency. Overvaluation is a positive difference between the real exchange rate and the equilibrium real exchange rate.  Since finance is a non-tradable good, the financial sector (in contrast to the manufacturing or agricultural sector) experiences growth during periods of real appreciation.  But financial institutions are also best positioned to take advantage of periods of overvaluation by shifting financial assets abroad into offshore accounts, something that domestic investors whose wealth is concentrated in non-financial assets are less likely to be able, willing, or accustomed to doing.  It is therefore problematic for the financial sector to be able to control the nominal exchange rate, allowing it to produce periods of overvaluation during which it can purchase cheap foreign currency and move it offshore. Venezuela is atypical among developing countries in that its accumulations of public external liabilities are typically mirrored more than one-for-one in accumulation of private external assets.}  This exchange-rate policy continued through democratization in 1958. Even during the heyday of import-substitution industrialization in Latin America, Venezuela’s industrial policy largely subsidized producers of non-tradables \citep{fajardoSectorFinancieroPublico1984} and involved tariffs lower than the regional average \citep{rodriguezCaudillosPoliticosBanqueros2004}. In short, the traditional problem of Dutch Disease—growth in the services sector and contraction of non-oil tradable production—was fostered rather than slowed by economic policy in Venezuela.\footnote{On Dutch Disease, see \citet{neary1986natural,lederman2007trade,lederman2012does}.}

Second, the private sector dominated oil production until nationalization of the industry in 1976 \citep{osmel2010political}. This private-sector dominance was in part the result of an external constraint: the United States sent a strong signal that, after oil nationalization in Bolivia (1937) and Mexico (1938), it would not tolerate another nationalization \citep{singhOilPoliticsVenezuela1989,tinkersalas2005}.\footnote{Indeed, Washington engineered a coup in Iran in 1953 to revert nationalization there.} Instead of nationalizing, authoritarian governments honored transferable oil concessions granted mainly during the rule of Juan Vicente Gómez (1908–1935); these concessions allowed owners to produce oil and pay a concession-specific royalty rate.  Successive military governments through Marcos Pérez Jimenez (1948–1958) maintained  a working relationship with multinational oil companies while introducing tax reforms that progressively raised the government’s take. After the transition to democracy in 1958, ruling party Acción Democrática simply raised the government’s effective fiscal take to 50 percent of profits\footnote{Governments prior to 1958 also claimed that the fiscal take was 50\%, but AD argued that the \emph{effective} take was lower.}---and made an explicit commitment not to grant or renew expiring concessions. Because most concessions were due to expire in the 1970s and 80s, oil giants saw the writing on the wall and investment declined (see Figure \ref{fig:oil}). Over the following fifteen years, successive democratic administrations progressively increased the tax take, raising oil taxes in 1958, 1966, and 1970 \citep{martinezPetroleoCrudo1997}. The oil industry was finally nationalized in 1976. Avoiding nationalization during the first half-century of oil production likely contributed both to economic growth during this period and to the subsequent successes of PDVSA, the state oil company.

Finally, Venezuela—like many other oil producers, and like many other Latin American countries—failed to save oil revenue during the 1970s and accumulated too much debt. By 1989, Venezuela's debt had ballooned to 62\% of GDP, from less than 7\% of GDP twenty years prior. Contrary to popular wisdom, most of that accumulation occurred not in the 1970s but during the 1980s, when debt rose from 17\% to 62\% of GDP. Like much of the rest of the region, as Bertola and Ocampo discuss in the introduction to this volume, Venezuela found it increasingly difficult to service this debt after the Federal Reserve raised interest rates in the early 1980s. What sets Venezuela apart is that it accumulated external debt \emph{despite} benefitting from one of the strongest positive terms of trade shock in the region. Venezuela's strongly procyclical spending makes it a canonical example of the ``voracity effect'' \citep{tornell1999voracity}, in which terms-of-trade windfalls lead to a more-than-proportionate fiscal response.

By the end of 1982, Venezuela had fallen into arrears on its external debt and entered negotiations to reschedule debt service.  This debt overhang contributed to slow growth throughout the 1980s.  Yet even after restructuring its obligations under the Brady plan in 1989, Venezuela had a hard time reestablishing growth.  Instead of the “lost decade” suffered by most of Latin America, Venezuela lost an entire quarter-century.

\section{The Lost Quarter-Century}

Between 1982 and 1999, Venezuela suffered the worst economic performance of any Latin American country except war-torn Nicaragua. Mismanagement of the oil boom in the 1970s had set Venezuela up for an inevitable recession when oil prices crashed. But the length and depth of the recession were by no means pre-ordained. Rather, Venezuela’s political institutions hamstrung the government’s response to the crash. In particular, the proliferation of veto players—instituted to constrain executive overreach—delayed and truncated adjustment that might have enabled economic recovery. This rigidity made it difficult to adjust to negative external shocks.

After nearly sixty years of authoritarian rule under Andean military dictators, Venezuela’s 1958 transition to democracy marked the beginning of more than half a century of competitive presidential elections. The relative stability of this system stemmed in part from a pact—the Pact of Puntofijo—among what were then Venezuela’s three largest political parties. The pact committed signatories to respect the outcome of elections, to share power, and to uphold certain common policy objectives. Venezuela’s dominant political party, Acción Democrática (AD), abandoned the idea of a Marxist-style vanguard in order to pursue broad and multi-class membership \citep{collier2002shaping}; other Venezuelan parties followed suit. 

The formal power-sharing agreement of the Puntofijo Pact was short-lived, but the culture of consensus-building and non-aggression between political actors partially survived.  The Puntofijo Pact is often likened to the National Front in Colombia, but they were actually quite different \citep{meucci2006pactos}. In Venezuela, there was no explicit commitment to alternation in power; moreover, the coalition government broke down in 1962, just four years in.\footnote{URD, one of the three parties to the original agreement, left the coalition government in 1962.  When Raúl Leoni (AD) assumed the presidency in 1963, he attempted to reassemble the coalition government (relabeled the Broad Base government), but he could not convince COPEI to join.  URD left the Broad Base government in 1968, marking the end of coalition governments. See Pacto de Puntofijo and Amplia Base, Fundación Empresas Polar (\href{fundacionempresaspolar.org}{fundacionempresaspolar.org}).} Divisions within AD created confusion about precisely which factions were parties to the pact. This flexibility may have made Puntofijismo more stable: as Fergusson and Vargas argue in this volume, the Colombian experience reveals that rigid power-sharing agreements can falter and spark violence when political clout shifts. 

These political institutions worked reasonably well in the 1960s and 1970s, when oil prices were high and rising (\citealp{dunning2008crude}, though see \citealp{myers1975urban,McCoyMyers,dietz2007thaw}). But they were ill-equipped to handle the double shock of the 1980s: the collapse of oil prices and the debt crisis \citep[see e.g.][]{coppedge1997strong}. These shocks affected other Latin American countries, too. But Venezuela’s resultant recession was the worst in the region.
Certainly, as the literature has documented extensively, part of this outcome was the result of policy decisions made during the oil boom of the 1970s. For one thing, Venezuela’s only export was oil. Had there been other exports, the exchange-rate depreciation caused by falling oil prices could have fueled an expansion in those other sectors; in their absence, the newly depreciated exchange rate was insufficient to make Venezuela competitive in anything else \citep{rodriguezWhyResourceabundantEconomies1999,hausmannVenezuelaAnatomyCollapse2011}. Moreover, even sectors producing non-tradables were intensive in imported intermediate inputs and capital goods; therefore, a decline in oil revenues would also shrink non-oil output, amplifying the contractionary effect of an adverse terms-of-trade shock \citep[chap.~4]{rodriguez2021a}. 

Yet even from this weak starting point, Venezuela underperformed. Throughout the 1980s and 1990s, politicians initiated the necessary reforms—only to be blocked by one or more of the many veto players built into the country’s democratic institutions.

Consider, for example, the challenge of devaluation. When Luis Herrera Campins assumed the presidency in 1979, the Venezuelan economy was overheated and over indebted. Of special concern was the current account deficit, which reached 15\% of GDP the previous year. The fiscal deficit, in contrast, was moderate—2.7\% of GDP—thanks to significant spending cuts in the final year of the previous administration \citep{rodriguez2012,lopezobregon2001}. To address the current account deficit, Herrera needed either to increase the price of imported goods or to decrease the price of domestic goods. He could have accomplished the former by devaluing the overvalued exchange rate. To accomplish the latter without devaluing the exchange rate, he would have had to slash government spending and commit to contractionary monetary policy. Instead, Herrera slashed government spending \emph{without} credibly committing to contractionary monetary policy, thereby engineering a recession while maintaining an overvalued exchange rate.  The result was massive capital flight: the private sector accumulated \$15bn in net external assets between 1980 and 1983, equivalent to nearly half of the country’s outstanding external debt (\$33bn in 1984) \citep{mayobreRenegotiationVenezuelaForeign2019}.

 Why did an administration that had shown its willingness to engage in costly adjustment—even going so far as to engineer a recession in the midst of rising oil prices—so fervently resist devaluation, which would almost certainly have been less contractionary?  Why did it go on accumulating massive external liabilities despite recognizing the need to curb spending? Why did it effectively use public borrowing to subsidize the private sector’s accumulation of external assets?  Worse still, why did Herrera later adopt a multi-tier exchange rate in which certain importers and certain private-sector holders of external debt could buy foreign currency at a preferential rate?
 
The answer to all of these questions lies in intra-elite conflict. The nineteenth-century accommodation between Liberals and the commercial–financial sector began a long tradition of appointing representatives of the banking sector to key economic policy positions. Herrera’s finance minister was no exception. Arturo Sosa was a former banker with close ties to the Vollmer Group, a conglomerate with interests in banking and manufacturing; he had also served as finance minister for the civilian--military junta that took power after the fall of the dictatorship in 1958. And it was Sosa who pushed for a multi-tier exchange rate rather than devaluation with a unified rate, allowing cronies in the private sector to use cheap dollars to repay dollar-denominated debt. Sosa later likened this policy to an act of justice, saying that failing to guarantee the preferential exchange rate for private-sector debt repayment would have constituted “a mere asset confiscation” \citep{beroes1990recadi,vasquez2021}. This soft institutionalization of the power of the financial sector created a veto player that blocked exchange-rate adjustments over and over again. In just three weeks before the Herrera administration established the multi-tier exchange rate in 1983—a partial devaluation—the private sector moved \$890 million dollars out of the country: 1.3\% of that year’s GDP (Henry, 2003, p. 103). 

The influence of the financial sector was felt far beyond exchange-rate policy. The banking lobby also defeated several attempts at banking reform.  By the late 1980s, it was clear that the oligopolistic structure of the country’s financial sector was a problem: in 1989, seven banks held 63\% of all deposits. Ownership of those banks was highly concentrated, typically in the hands of few families, which affected corporate governance. This allowed bankers to focus lending on affiliates and use that focus to exert political influence. Moreover, banking regulations at the time isolated the industry from the rest of the world, creating uncompetitive structures and extraordinary profits. When the administration of Carlos Andrés Pérez (1989--1993) tried to change the rules of the game with a new Banking Act, co-written with the World Bank, that formed part of his sweeping reform program, bankers managed to delay. They had bankrolled Pérez's campaign and his political party at large, and they began to condition further support on the party’s opposition to reforms. In the end, the new act implied only minor changes for the banking industry \citep{dekrivoy2000}.  

The financial sector was not Venezuela’s only veto player.  The powerful federation of labor unions, for example, blocked overdue labor-market liberalization until 1990. As the economy entered recession, labor market restrictions became even more onerous, swelling the informal sector and dampening labor productivity. For the same reason, Venezuela’s fiscal adjustment in the 1980s was strongly biased against long-term investment in public goods such as infrastructure, generating additional adverse productivity effects \citep{pineda2014public,moreno20219}. Political parties in Congress also blocked a critical tax reform by the Pérez administration that would have introduced a value-added tax \citep{naimPaperTigersMinotaurs2013}.  The same congress then approved the reform after Pérez was impeached—but not without extracting significant increases in transfers to regional governments in exchange \citep{riveroRebelionNaufragos2016,pineda2014public,corrales2010presidents}. 

The intra-class nature of political conflict in Venezuela in the 1980s and 1990s is reflected in voting patterns. \citet{kronick2021backsliding} show that, in line with a large body of qualitative work \citep[especially][]{handlin2017state} but in contrast to influential theory papers, voting for Chávez in 1998 split \emph{across} class lines, not \emph{along} class lines. The elite factions that faced off in the mid-90s banking crisis, for example, went on to support different candidates in 1998 \citep{gates2010electing}. Chávez came to power as a populist in the sense of \citet{barr2009populists}: peddling an anti-system message that appealed to voters from across the income spectrum \citep[see also][]{hawkins2010venezuela}; only later did he become a macroeconomic populist in the sense of \citet{edwards1991macroeconomic,acemoglu2013political}. This is not to deny the role of class conflict during this period---only to stress that intra-class conflict clearly contributed to Venezuela's twentieth-century economic collapse and to the rise of Chavismo. 

%
%

\section{The Worst Recession} 

Like the recession of the 1980s, Venezuela’s most recent recession (2012–2020) was due in part to the unraveling of a standard macroeconomic populist cycle \citep{edwards1991macroeconomic}. Rather than save or invest the oil windfall of 2003–2012, the government squandered it on a tremendous consumption boom, leaving Venezuela unprepared for a large drop in oil prices. But the magnitude of this most recent recession cannot be explained by the sins of previous administrations alone. In Venezuela’s third collapse, per capita GDP contracted by 71.8\%—more than in any other recession in Venezuelan history, and more than any other recession in Latin America since the advent of comparable cross-national data in 1950. It is difficult to fully explain this collapse without considering the role of intra-elite conflict, which, once again, created incentives for policy that severely damaged the economy.

\paragraph{A large part of Venezuela’s 2012–2020 recession} stemmed from economic policies enacted during the preceding oil boom. Hugo Chávez, elected president in 1998, eventually rolled back most of the market-oriented economic reforms of the 1990s \citep{corrales2011dragon}. He re-instituted exchange controls; expanded and tightened price controls; nationalized hundreds of firms, including major companies in telecommunications, oil services, and banking; and even imposed a 19-year firing freeze. 

\begin{figure}[t!] 
\RawFloats
       \centering
       \captionsetup{width=1\textwidth}
       \caption{Venezuelan Oil Production, 1920--2020}\label{fig:oil}
      \caption*{\footnotesize{This figure plots Venezuelan oil production from 1920 through 2020. Production fell in anticipation of nationalization (1976), increased in the 1990s with the \emph{apertura}, dipped during the oil strike of 2002–2003, declined slowly under Chávez, and collapsed beginning in 2016. The inset shows the post-2016 period.}} \vspace{-.01cm}
 \label{fig:mainresults}
  \begin{tikzonimage}[width=1\textwidth, trim= 0cm 0cm 0cm .5cm, clip=true]{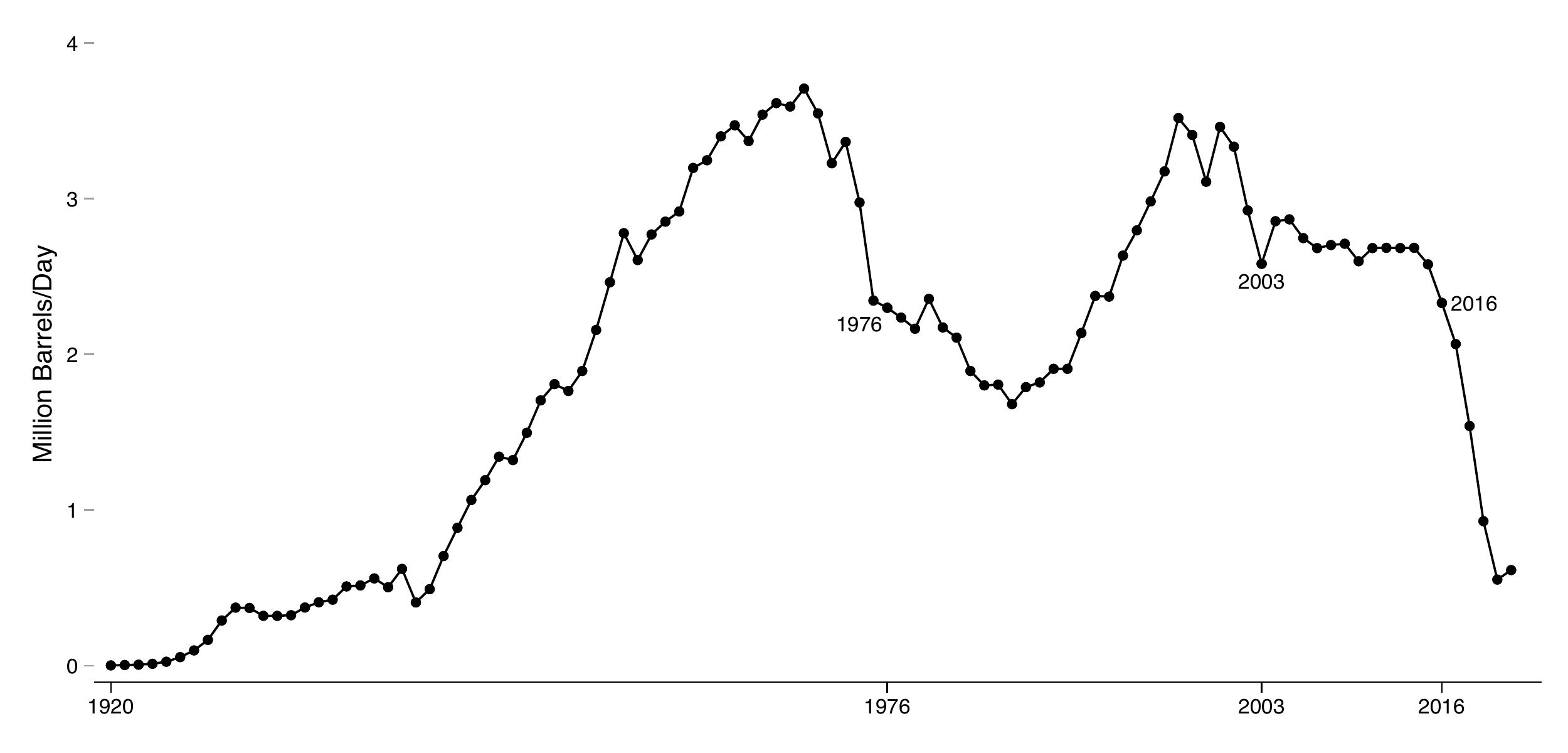}
\node at (0.335,-0.35) {\includegraphics[width=0.5\textwidth]{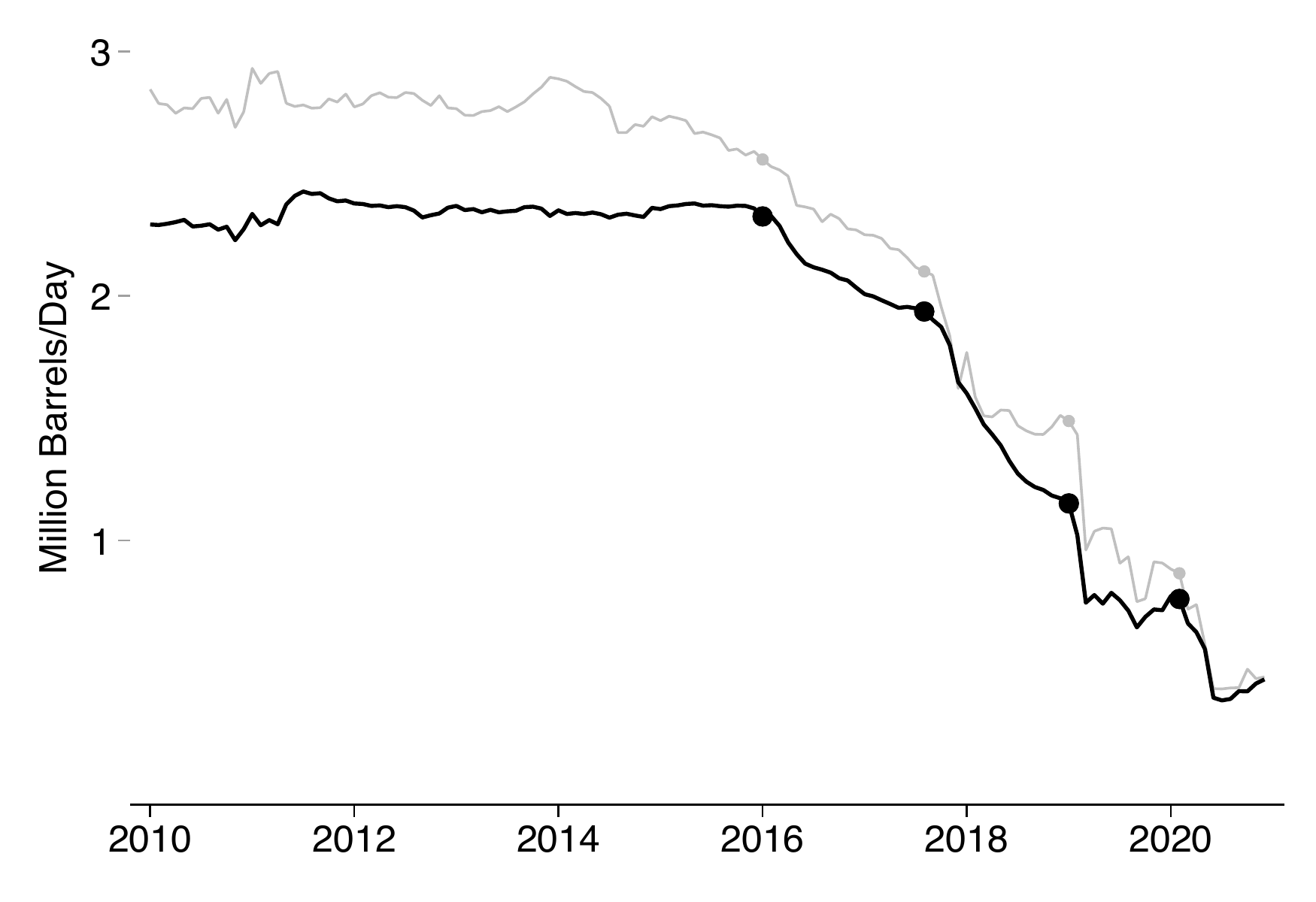}};
   \draw [draw=red,opacity=0.25,line width=2pt] (0.865,.725) rectangle (0.975,0.2);
   \draw [draw=red,opacity=0.25,line width=2pt] (0.085,0.025) rectangle (0.585,-0.72);
   \draw[->, dashed, color = black, line width = .5pt] (0.61,-0.15) to[out=180,in=0] node {} (.41,-0.15); 
   \draw[->, dashed, color = black, line width = .5pt] (0.61,-0.23) to[out=180,in=0] node {} (.46,-0.23); 
   \draw[->, dashed, color = black, line width = .5pt] (0.61,-0.39) to[out=180,in=0] node {} (.51,-0.39); 
      \draw[->, dashed, color = black, line width = .5pt] (0.61,-0.47) to[out=180,in=0] node {} (.555,-0.47); 
\node [anchor=west,  text = black, font=\scriptsize\fontfamily{phv}] at (0.61,-0.15) {01/2016: Oil price falls below \$30/barrel};
 \node [anchor=west,  text = black, font=\scriptsize\fontfamily{phv}] at (0.61,-0.23) {08/2017: U.S.\ imposes financial sanctions};
  \node [anchor=west,  text = black, font=\scriptsize\fontfamily{phv}] at (0.61,-0.39) {01/2019: U.S.\ imposes primary oil sanctions};
    \node [anchor=west,  text = black, font=\scriptsize\fontfamily{phv}] at (0.61,-0.495) {\specialLcell{02/2020: U.S.\ imposes secondary oil \\ sanctions on trading partners}};
 \end{tikzonimage}
\caption*{\scriptsize Sources: \citet{baptista1997bases}, EIA. The gray line in the inset marks Venezuelan oil production as reported by the government (to OPEC); the black line plots production as reported by independent agencies to OPEC. We include both series because the government data show a decline in 2010--2015; even with this pretrend, however, the post-sanctions declines appear dramatic.}
 \end{figure}

Chávez’s treatment of the oil sector resulted in underinvestment, mismanagement, and disappointing production \citep{monaldi2018death,monaldi2021collapse}. He progressively raised taxes on oil companies, revising hydrocarbons legislation in 2001, 2004, 2006, and 2008. At the same time, he forced multinationals to convert their investments into joint ventures with the state-owned PDVSA. He stacked the PDVSA board with political loyalists and fired nearly half of company employees \citep{hsieh2011price,corrales2011dragon}, replacing them with new hires who, in many cases, lacked expertise. The predictable result was that Venezuela's oil industry underperformed, relative to peers. Oil production actually declined between 1998 and 2008 (Figure \ref{fig:oil}), despite the fact that Venezuela had the largest oil reserves in the world (296.5bn barrels, according to a 2010 PDVSA certification), and despite soaring oil prices. In the midst of this tremendous windfall---by far the largest positive terms-of-trade shock in Latin America during this period \citep{kronick538}---  Chávez committed the classic macroeconomic populist mistakes of accumulating debt, failing to build up reserves, and running fiscal deficits. By 2012, the public-sector deficit hit 17\% of GDP.

Some scholars view these destructive policies as the product of class conflict: Chávez sought to redirect oil revenues toward his constituents, while traditional elites fought to keep the money for themselves. There is of course some truth in this interpretation, but, in our view, intra-elite conflict deserves more emphasis. Chávez created two off-budget funds—the National Development Fund and the Chinese Fund—that were under discretionary control of the executive. Contrary to the government’s claims, social spending (as a share of fiscal spending) did not systematically increase, the much-touted social programs or \emph{missions} failed to deliver consistent results \citep{rodriguez2008empty,ortega2008freed}, and the subsidized official exchange rate did at least as much to line the pockets of elites as it did to lower prices for poor consumers \citep{gulotty2022arbitrage}. Nor did Chávez deliver on his campaign promise of improving citizen security \citep{kronick2020profits}. Meanwhile, government cronies racked up tens billions of dollars in private wealth \citep{Pandora}. In that sense, this period is best modeled as a clash between two powerful groups.

How Chávez amassed the power to implement these policies has been well studied. \citet{corrales2018fixing} shows how Chávez's wide margin of victory in the 1998 election paved the way for a new constitution that empowered the president; \citet{brewer2010dismantling} documents how Chávez pulled off subsequent power grabs, including, critically, attacks on the media \citep{knight2019limits,knight2022opposition}. A newer literature, building on \citet{corrales2005search}, considers the role of the opposition \citep{jimenez2021contesting,velasco2022many}, finding that certain opposition tactics proved self-sabotaging, strengthening Chávez's hand \citep{gamboa2017opposition}. For our purposes, what is critical is that Chávez-era economic policies left Venezuela ill-prepared to weather the next drop in oil prices, which began in 2014.


\paragraph{Chávez died of cancer in March, 2013,} anointing Vice President Nicolás Maduro as his successor.  One month later, the government held a snap election to determine who would complete Chávez’s term. Maduro defeated the opposition candidate, former governor Henrique Capriles, by a margin of just 1.2 percentage points.  Upon taking office, Maduro faced an urgent economic-policy challenge. Prior to the 2012 presidential election—which Chávez won with a comfortable margin—the government had embarked on a spending spree, raising public spending by more than 10 percent of GDP  \citep{rodriguez2021a}. The result was a hefty budget deficit. It fell to Maduro, then, to drastically cut spending. 

This adjustment would have been painful no matter how Maduro carried it out. But the path he chose was much more destructive, in the medium term, than alternatives that the government explicitly considered. One standard approach would have been to devalue the currency and eliminate a lavish fuel subsidy, accepting significant short-term inflation in return for improved public finances and the potential for growth in non-oil export sectors \citep{ToroKronick}. Instead, Maduro maintained a massively overvalued exchange rate, draconian price controls, and the fuel subsidy. He paid for all of this with a dramatic cut in imports, which fell 28\% between 2012 and 2014. In short, Maduro chose to adjust via quantities rather than via prices—creating major shortages and distortions, exacerbating fiscal imbalances, and ultimately deepening Venezuela’s economic contraction.

When oil prices crashed in 2014, creating the need for more spending cuts, Maduro doubled down on the previous year’s policies: slashing imports in order to maintain expensive subsidies for foreign currency and domestic fuel. To these measures he added inflationary financing. The result was a 37\% contraction in his first four years in office (March 2013–March 2017). Even with the oil-price crash, the recession would almost certainly have been milder if Maduro had corrected misalignments from the beginning. 

Some argue that Maduro chose this destructive adjustment path because of corruption, incompetence, or ideology. These explanations are incomplete. Funneling money to cronies does not require destroying the economy \citep{coate1995form}. And the fact that Maduro repeatedly announced the intention to devalue, to lift price controls, or to otherwise liberalize in ways that would get the economy back on track---as well as the fact that he ultimately did implement many of these reforms when he stopped facing electoral challenges after 2018---reveals that doing so was neither ideological anathema nor beyond the government’s competence. \citet{rodriguez2021a} argues instead that Maduro’s choices were motivated by Venezuela’s relentless electoral cycle. Nationwide elections were held in three of his first five years in office (2013, 2015, and 2017); in the other two years (2014 and 2016), he faced sustained nationwide protests, and (in 2016) the possibility of a recall referendum. In this environment of constant campaigning, shifting costs into the future was more desirable than accepting short-term costs in return for medium-term gains. 

Of course, Venezuela is not the only country to experience a political business cycle in which policies grow more distortionary as elections approach \citep{rogoff1988elections,alesina1997political}. But, as \citet{Zimbabwe} argue, the combination of high-frequency contested elections and high stakes of power in competitive authoritarian regimes makes this cycle especially pernicious. Both Zimbabwe and Venezuela---which experienced two of the largest peacetime growth collapses since 1950, and which suffered hyperinflation in the twenty-first century---adopted more distortionary policies immediately after significant opposition electoral victories. Contested elections for absolute power led both sides to adopt strategies with tremendous economic externalities.

Most of the distortionary policies were, of course, chosen by the government, as we have described. But the opposition also played a role. Under Maduro, lobbying for financial and trade sanctions became a crucial part of the opposition’s strategy. Indeed, \citet{esberg2021exile} find that opposition activists who leave Venezuela tweet about sanctions (and even foreign military intervention) twice as often as their peers who remain in the country. These sanctions damaged the country’s oil sector. In this sense, Venezuela’s post-2012 growth collapse is a combination  of two distinct episodes: in 2012--2016, a ``classic 80s crisis'' \citep{kronick2020} that stemmed from accumulated imbalances (i.e., prior overspending and undersaving), the oil-price shock, and Maduro’s decision to maintain relative price imbalances; and, in 2017--2020, a crisis driven largely by the decline of oil production---which, in turn, stemmed in part from financial and trade sanctions imposed by the United States.

To see this, consider that oil production had remained relatively stable between 2008 and 2015, after the government had agreed on a modus vivendi with private sector partners (Figure \ref{fig:oil}). It began to decline in early 2016, when oil prices dipped below \$30/barrel, and then declined sharply in August 2017—when the United States imposed financial sanctions—and again in January 2019, when the United States imposed oil sanctions (see Figure \ref{fig:oil}, inset) \citep{monaldi2018death}. In 2020, when the U.S. imposed secondary sanctions on Russian and Mexican companies helping market Venezuelan oil, production declined sharply yet again \citep{rodriguezSanctionsVenezuelanEconomy2019,rodriguez2021a}. And while any single time series admits multiple interpretations, there are clear mechanisms linking sanctions to oil production. For example, data from joint ventures between state-owned PDVSA and multinationals operating in the country’s Orinoco Basin shows that firms with access to foreign financing through their partners suffered a much stronger decline in production after financial sanctions than those that lacked such access \citep{rodriguez2021b}.
	
	These costly sanctions are, of course, primarily a response to Venezuela’s Maduro-led transition to autocracy. After losing control of the legislature in 2015, Maduro used his control of the judiciary to strip the National Assembly of its powers, invalidate the 2016 drive to hold a recall referendum \citep{eubank2021friends}, and ban several competing parties and candidates from the 2018 presidential election. These abuses and others have been chronicled in detail elsewhere \citep{corrales2020authoritarian,bahar2018venezuela}. But the sanctions are also a result of advocacy by the political opposition, aimed at blocking the Maduro administration’s access to resources.  This effort intensified after 2017, when the opposition-controlled National Assembly publicly called on foreign financial institutions to abstain from approving financing to the Maduro government.  The opposition strongly supported U.S. sanctions on the Venezuelan oil company, which factored into the U.S. decision to adopt them \citep[p.~250]{bolton2020room}.
	
Why were politicians willing to engage in modes of political conflict that were so costly for the economy? Part of the answer is that the stakes of political power were so high \citep{przeworski1991democracy,monaldi2006political}. In his first year in office, Hugo Chávez had raised the stakes of power through a new Constitution (1999) that granted expansive new powers to the executive: running for immediate reelection, controlling promotions in the military, calling popular referenda, and even convening new constitutional conventions that could dissolve other powers of government \citep{corrales2011dragon,trinkunasReemergenceVenezuelanArmed2002}. These new, higher stakes of power generated economically destructive political conflict even while Chávez was in office \citep{corrales2005search}. Under Maduro, as \citet{rodriguez2021a} argues, high stakes of power combined with a negative economic shock to further increase the incentives for political strategies that destroyed the means of production.

\section{Political Conflict and Economic Growth}

The overarching fact about the Venezuelan economy since independence is that it dramatically underperformed during three periods: first in the nineteenth century, then in the late twentieth century, and most recently in the second decade of the twenty-first century. All three incidents dwarf typical recessions. In the nineteenth century, Venezuela failed to grow despite improving terms of trade. In the twentieth century, the Venezuelan economy shrank 26.1\%. And in the most recent and most extreme incident, the economy shrank by more than 70\%, making this the second-largest peacetime economic collapse in the world since 1950.

Certainly, dependence on natural-resource exports has made Venezuela especially vulnerable to these swings. And certainly, the latter two collapses are due in part to the unraveling of standard macroeconomic populist policy cycles \citep{edwards1991macroeconomic}. But we argue that the surprising magnitude of these catastrophes can only be explained by understanding the structure of intra-elite conflict. In the nineteenth century, in the absence of a federal government, landowners and financiers engaged in fighting that physically destroyed plantations and cattle ranches. In the twentieth century, new- and old-guard elites faced off in pitched political battles that hampered adjustment to the negative oil-price shocks. And in the twenty-first century, high stakes of power drew politicians into a destructive contest for international bank accounts and export receipts. These conflicts are critical to understanding growth performance. 




\newpage

\normalsize

\def\UrlBreaks{\do\/\do-}

\bibliographystyle{plainnat}
\bibliography{Library_HistoryChapter.bib}

\end{document}